\documentclass[12pt]{amsart}
\usepackage{amssymb}
\usepackage{ljm-auth}

\usepackage[T2A]{fontenc}
\usepackage[english]{babel}
\usepackage[utf8]{inputenc}

\usepackage{graphicx}

\tit{Quantum-Assisted Blockchain}
\shorttit{Quantum-Assisted Blockchain}

\author{D. Sapaev$^{1,3}$, D. Bulychkov$^{2,3}$, F. Ablayev$^{3}$, A. Vasiliev$^{3}$, M. Ziatdinov$^{3}$}
\crauthor{Sapaev, Bulychkov, Ablayev, Vasiliev, Ziatdinov}

\begin{document}

\maketit

\address{$^1$Sberbank-Technologies, Development department of Center of Technological Innovations,
Universitetskaya 7, Innopolis, 420500, Russia\\
$^2$Sberbank, Center of Technological Innovations, Vavilova 19, Moskow, 117312, Russia\\
$^3$Kazan (Volga Region) Federal University,
Institute of Computational Mathematics and Information Technologies,
35 Kremlevskaya str., Kazan, 420008, Russia
}

\email{d.a.sapaev@gmail.com}

\abstract{
Bitcoin and blockchain in general is a hot topic nowadays. In the paper we propose a quantum
empowering of this technology and show how to speed-up the mining procedure using the modified
Grover's algorithm.
}

\notes{0}{
}

\section{Introduction}

Quantum computing is a hot modern topic of Theoretical Computer Science. Theoretical research in
this area started in the last decades of previous century.
  Yuri Manin and Richard Feynman were one of the first famous researchers who
  proposed this area of research in 1980s. Recent results in quantum computations and
  quantum technology achievements brought theoretical results to the practice.
  The appearance of so-called IBM-Q device created a new quantum computer science
  community and gave it a tool to verify the known theoretical ideas and algorithms.

Another hot area in applied computer science is the blockchain technology. Clearly, that it will be
interesting to check out what opportunities can  quantum computation theory and quantum technology
give to the blockchain technology. For instance, there were several proposals on empowering Bitcoin
Electronic Cash System with quantum technologies (see, e.g.
\cite{Fedorov-et-al:2017:Quantum-secured-blockchain,Jogenfors:2016:Quantum-Bitcoin,Ikeda:2017:qBitcoin}),
as well as on possible attacks on this system
\cite{Aggarwal-et-al:2017:Quantum-attacks-on-Bitcoin}.

In this paper we consider the natural idea of applying Grover's quantum search algorithm to the
general blockchain technology (for mining). So, the paper is devoted to developing and testing this
idea for the blockchain technology.

\section{Preliminaries}

In the classical case, the complexity of searching in an unstructured data set of size $n$ is
\(O(n)\), e.g., in the worst case, we have to look through all the records. The well-known Grover's
algorithm \cite{Grover:1996:Fast-Search} allows to solve this problem in \(O(\sqrt{n})\) steps.
Thus, if we have 40 bits and we need to find a combination that satisfies certain condition, then
in the classical case we need to process about \(1\,000\,000\,000\,000\) different combinations,
whereas the quantum algorithm will yield a result in about  \(1\,000\,000\) queries. We briefly
recall the structure of the algorithm.

We have a register of \(n\) qubits that are in the uniform superposition of the basis states, i.e.
we have all possible \(2^n\) states, the amplitude of each state is \(1 / \sqrt{2^n}\) (the sum of
the squares of the modules will give 1).

We also have one additional qubit, which we will call functional, and the function is an oracle.
The oracle translates the functional qubit from the state \({\left\vert{0}\right\rangle}\) to the
state \({\left\vert{1}\right\rangle}\), if all the other qubits state encodes the item that we are
looking for.

In order to get exactly the state that corresponds to the oracle function, we start with changing
the sign of the amplitude of the desired value to negative (to do this we translate the functional
qubit into the state \({\left\vert{1}\right\rangle}\) on all sets, then apply the Hadamard gate to
it and apply the oracle function to the whole register).

After the sign of the target amplitude has changed, the average value has dropped below the average
value of the amplitudes. 

The next step is the Grover diffusion operator, that makes the value of the sought-for state's
amplitude higher than to the others. Performing this procedure about \(O(\sqrt{2^n})\) times will
make the desired amplitude close to unity.

\section{Quantum-Assisted Blockchain}

Since quantum computers can perform an exhaustive search quadratically faster than classical
computers, we can use the modified Grover's algorithms to perform mining on a quantum computer.
Indeed, if we can consider all the values of nonce at once, then we can speed up the search for the
right one.

To implement this idea, we start with dividing our quantum register into several parts: nonce (we
apply the Hadamard transform to these qubits and we will consider all possible values at once),
hash (there will be corresponding hash values), service qubits (for implementing basic operations)
and a functional qubit for the Grover's algorithm.

The algorithm itself consists of the following steps:
\begin{enumerate}
\item We apply the Hadamard transform to the nonce qubits and calculate
the hash value for all of the nonce values at once (we need to implement a quantum procedure
for the SHA-3 hashing algorithm here);
\item For each block of the incoming message (block header in the blockchain),
we first mix it with the hash state (in the SHA-3 terms), and then compute the 
hash function. We get a register that contains all values of nonce, hash values for each nonce,
a number of service qubits that are needed to temporarily store the intermediate computations
and a functional qubit;
\item We use the oracle function to calculate a hash value that is below a certain threshold.
This function is a NOT operation controlled by those qubits whose value is intended to be zero
in the desired hash value;
\item Apply the Grover's algorithm to find the desired hash value and nonce.
\end{enumerate}

\subsection{Implementation of logical operations using quantum transformations}

To implement classical hashing algorithms on a quantum computer, it is sufficient to show how to
implement the following set of primitives -- XOR, AND, NOT and bitwise shift.

\textbf{XOR} is implemented using the CNOT gate. But in the case of a hash, we would not like to
write the result to one of the operands, thereby erasing its value. To write the result of an XOR
operation into a separate qubit, one have to:
 \begin{enumerate}
\item Initialize the service qubit in the state \({\left\vert{0}\right\rangle}\);
\item Perform a CNOT gate, in which the first operand is the controlling one, and the service qubit
is a target;
\item Perform the same transformation, but with the second operand as the
controlling one.
\end{enumerate}

Thus, the service qubit will be in the state \({\left\vert{1}\right\rangle}\) if and only if
exactly one of the operands is 1, otherwise it will be \({\left\vert{0}\right\rangle}\). And this
is exactly the XOR operation.

\textbf{AND} is implemented using the three-bit gate CCNOT -- it inverts the target qubit only when
the first two are in state \({\left\vert{1}\right\rangle}\).

\textbf{NOT} is implemented by a simple Pauli gate $X$.

\textbf{Bit shift} can be implemented using series of swap transformations (qubit exchange).

\subsection{Problems of using Grover algorithm for mining}

The proposed algorithm has to deal with the following issues.

\paragraph{Too low value of the average.}
Grover's algorithm works efficiently only if we have a uniform superposition of all the qubits
participating in it. But in our case we put only the first \(n\) qubits of nonce into the
superposition, and the remaining \(m\) qubits contain the corresponding values of the hash
function. This means that instead of \(2^{n+m}\) different states, we have only \(2^n\), the
remaining states have zero amplitudes. This means that the average value is not close to the
amplitude \((1 / \sqrt{ 2^n}\), but closer to $0$, because when calculating the average, the sum of
all the amplitudes is divided by the total number of states.

Therefore, after the first Grover diffusion operator all amplitudes go to negative values, which
does not allow Grover's algorithm to be used further.

\paragraph{The increase in the zero amplitudes.}
If we consider not the entire quantum register, but only the block that corresponds to a certain
value of nonce, then we see an incomplete distribution -- only one amplitude is nonzero (it
corresponds to the value of the hash function with this nonce value), the other amplitudes will be
0. Full distribution for nonce \(00 \ldots 0\):
\[
a_0 {\left\vert{{00\ldots 0 \, 00\ldots 0}}\right\rangle} + a_1 {\left\vert{{00\ldots 0 \, 00\ldots
1}}\right\rangle} + \cdots + a_m {\left\vert{{00\ldots 0 \, 11\ldots 1}}\right\rangle} .
\]
Incomplete distribution for nonce \(00 \ldots 0\):
\[
0 {\left\vert{{00\ldots 0 \, 00\ldots 0}}\right\rangle} + 1 {\left\vert{{00\ldots 0 \, 00\ldots
1}}\right\rangle} + \cdots + 0 {\left\vert{{00\ldots 0 \, 11\ldots 1}}\right\rangle}.
\]
That is, the value of nonce \(00\ldots 0\) corresponds to the single value of hash -- \(00\ldots
1\), all other states can not be met.

This helps the oracle -- it ``marks'' a state with a non-zero amplitude only if it satisfies the
condition. The remaining states, even if they will satisfy the condition, will not be labeled,
since they have a zero amplitude.

But this advantage will vanish immediately after the first application of the Grover diffusion
operator -- we make the same transformations over all states simultaneously, and so the zero
amplitudes are also inverted and will be nonzero. 

\paragraph{The possible existence of many solutions.}
Grover's algorithm increases the amplitude only when applied certain number of times, afterwards
the reverse effect begins. This can be compared with the rotation of a vector -- it is rotated
several times by a certain angle and as soon as it passes the desired position, the vector starts
to move away.

We know the exact number of iterations of the Grover algorithm for the maximal success probability,
but this number is calculated for the case when the solution is unique. In our case, there can be
many such solutions, and therefore we do not know exactly how many times the Grover iteration
should be applied.


The main problems arise from the fact that if we consider the registers nonce and hash together,
then we have an incomplete distribution. To solve the first 2 problems, it is sufficient to switch
to the full register, and this can be done very easily -- by making the hash computation part of
the oracle.

Indeed, if we apply the Grover algorithm only to the nonce register, and the hash register is used
only for the oracle function to work, then at the end of the algorithm we can easily get the hash
value for the found nonce. This will just add one hash computation to the algorithm.


But it is impossible to implement the Grover algorithm only on the nonce register without
additional transformations. And the thing is that we entangle the qubits from the
nonce register with the hash register during the computation. 
And this means
that the simple application of the Grover algorithm in our case is impossible.

The solution to the problem is the inverting of all service qubits to the initial state. All of the
qubits that are not used in the Grover algorithm are to be transformed to the initial state --
\({\left\vert{0}\right\rangle}\). In our case, after evaluating the value of hash, the nonce
register becomes entangled with the hash register, which in turn is entangled with the service
register. All these connections must be broken each time before applying the Grover diffusion
operator.

Thus, the algorithm now has the following steps:
\begin{enumerate}
\item Transform the nonce register to the uniform superposition state;
\item Compute the hash value for all of the nonce values using quantum parallelism. The result will be in the hash
register;
\item Apply the oracle function to the hash register to find out which nonce values give the hash that is below
 a certain threshold;
\item Apply the reverse hash computation procedure to ``unwind'' all the qubits except for nonce and functional qubits to the initial basis state
\({\left\vert{0}\right\rangle}\);
\item Apply the Grover diffusion
operator to the nonce register;
\item Repeat steps 2--5 as many times as necessary.
\end{enumerate}

The solution to the third problem is in adjustment of the complexity of the network.

As mentioned above, the Grover algorithm is efficient if there is a unique solution that satisfies
the conditions. But the same problem arises also for the classic miner -- the PoW consensus assumes
the presence of a certain threshold for the value of hash. In the classical case, this allows us to
adjust the time of mining of one block.

In our case, this will reduce the probability of having more than one suitable value.

We have an \(n\)-qubit nonce register and an \(m\)-qubit hash register. For example, in bitcoin,
these values are 32 and 256, respectively. But since the power of a set of nonce is many times
smaller than the set of hash values, we can come up against the situation that after having gone
over all the values of nonce we did not get the hash, which satisfies our conditions. In such
cases, it is allowed to extend the time or change the transaction and try again. Therefore, without
loss of generality, we can extend the length of the nonce to 48 bits.

We expect the probability of a single correct solution to be close to $ 1/2 $, the total number of
solutions is \(2^{256}\), we divide it into \(2^{48}\) different sets. The probability that we have
one correct solution is the number of correct solutions divided by the number of different sets.
From the equation \(2^x / 2^{256-48} = 1/2\) we get that we must have \(2^{207}\) different
solutions, so the first 49 bits in the desired values of the hash should be 0. And this practically
coincides with the current level of complexity of the bitcoin network! Hence, we need 48 qubits for
nonce and 256 qubits for the hash value to implement the complexity threshold of the bitcoin
network.

\subsection{Comparative time analysis of the speed of the classical and quantum mining algorithms}

According to Intel's measurements, the average four-core classic computer at 3 GHz processes about
7 million hashing operations per second.

According to measurements, one gate per quantum computer requires about 1 nanosecond. And this
is 1 billion gates per second. 

To handle our 48-bit nonce, the classic computer will have to execute about \(2^{48}\) hashing
operations. Given the speed of 7 million operations per second, we get an estimate of the required
time -- about 40 million seconds, which is 11,000 hours or 465 days. It roughly converges with the
current reality of bitcoin -- all network majors in the world find the right value in 10 minutes.

A quantum algorithm requires \(\pi \sqrt{2^n} / 4\) operations. This is about 13 million
operations, and this, considering the same processing speed, is only 2 seconds!

\bibliographystyle{unsrt}
\bibliography{references}

\end{document}